\documentclass{article}[12pt]

\usepackage[dvips]{epsfig}
\usepackage{amsmath}

\begin{document}

\title{\textbf{Primordial bubbles of colour superconducting quark matter}}
\author{Luis Masperi$\dagger $\thanks{%
On leave of absence from Centro At\'{o}mico Bariloche, San Carlos de
Bariloche, Argentina.} and Milva Orsaria$\ddagger $ \\
$\dagger $ Centro Latinoamericano de F\'{\i}sica,\\ Av. Venceslau
Braz 71 Fundos, 22290-140 Rio de Janeiro, Brazil\\ $\ddagger $
Centro Brasileiro de Pesquisas F\'{\i}sicas, \\ Dr. Xavier Sigaud
150, 22290-180 Rio de Janeiro, Brazil}
\date{}
\maketitle

\begin{abstract}
We analyze the possibility that bubbles of quark matter surviving the
confinement phase transition might have become colour superconducting due to
the enormous compression suffered by them.

Because of the relatively high temperature of the process we compare the
initially unpaired quark matter with the colour-flavour locked alternative
when the extremely large chemical potential could have increased the
critical temperature sufficiently and find that this latter phase would be
more stable before the bubble compression stops.

If other physical effects have not affected completely their stability,
these bubbles might still exist today and perhaps be observed as strangelets.
\end{abstract}

\section{Introduction}

\noindent New phases of $QCD$ are being intensively studied with the
possibility of future phenomenological evidence both from accelerators and
astrophysical observations.

Apart from the ordinary hadronic phase and the expected deconfined
quark-gluon plasma ($QGP$) at high temperature, interesting colour
superconducting states of quark matter are theoretically predicted at very
high pressure$\cite{1}\cite{2}$.

There is agreement that the ultimate stable phase at extremely large
chemical potential $\mu $, and consequently pressure, is the colour flavour
locked one ($CFL$) where couples with zero total momentum of quarks of
different colours and flavours $u$ $d$ $s$ are paired due to the fact that
mass $m_{s}$ is no longer relevant. For decreasing chemical potential,
before reaching the ordinary nuclear matter, it is possible that
intermediate phases prevail such as colour crystals\cite{3} due to pairing
of quarks of different momentum, or Cooper pairs of only light quarks $u$ $d$
($2SC$), or even single flavour diquark condensates\cite{4}.

During the cooling of universe, when the temperature reached $T\sim 150\;MeV$
presumably a first-order phase transition occurred\cite{5}, even though
there is debate on this point, coincident with breaking of the approximate
chiral symmetry, passing from deconfined almost free quarks in thermodynamic
equilibrium to quarks confined in hadrons.

If the transition was of first order, it was produced by the expansion of
bubbles of low-temperature phase in the sea of the high-temperature one. But
when the coalescence velocity became smaller than that of the expansion of
universe, a reversed situation occurred, with bubbles of the surviving quark
matter surrounded by the more stable hadronic phase\cite{6}.

The negative pressure due to the energy of the false vacuum\cite{7} inside
these bubbles makes them contract. The chemical potential which remained
constant in both phases during the transition begins to increase inside the
bubble assuming no emission of baryons. This increase is not very relevant
during the first stage of the compression when quarks and antiquarks
annihilate with emission of neutrinos because of their very long mean free
path. But in the final stage, when almost all antiquarks have disappeared
due to the matter-antimatter asymmetry generated by earlier processes in the
universe, the chemical potential increase makes the quark pressure rise
rapidly and the bubble mass starts to increase too. Therefore the total
pressure becomes positive and tends to decrease the contraction velocity.

The interval of time of the bubble contraction turns out to be small
compared to the age $\sim 10^{-5}\sec $ of the universe at the confinement
transition. Therefore the temperature may be considered constant and must be
compared with the critical temperature for the loss of colour
superconductivity. For $T\sim 150\;MeV$ of all the superconducting phases
only the $CFL$ one can be probably reached at extremely high chemical
potential. This occurs when the bubble has not yet stopped its contraction
and from this moment the $CFL$ free energy for variable number of particles
can be evaluated as the sum of a free-quark contribution with the constraint
of equal density for the $3$ flavours, which is slightly larger than the
unpaired value, minus a term due to Cooper pairs gap which makes the
superconducting state more stable. Since this free energy $\frac{J}{V}=-p$,
the pressure increases even faster and the bubble contraction is very
quickly stopped.

In Sec.2 we give the thermodynamic expressions for massless and massive
quarks of flavour and energy densities and of pressure as functions of
temperature and chemical potential, including also a description of the $CFL$
phase. Sec.3 is devoted to the calculation of the bubble contraction through
the evolution of chemical potentials which determine its mass, pressure and
velocity. Finally in Sec.4 we comment on physical aspects which should be
also taken into account and on the possibility that a part of these bubbles
might have survived till our times and could be detected by high-altitude
observatories.

\bigskip

\section{Thermodynamics of free and paired quarks}

For the deconfined $QGP$ phase the high $T$ allows the approximation of
neglecting the interaction among quarks so that for each fermion of mass $m$%
, with $2$ spin projections and $3$ colours, the number and energy densities
and the pressure are given by
\begin{equation}
\frac{N}{V}=\frac{3}{\pi ^{2}}\int_{m}^{\infty }d\varepsilon \;\varepsilon \;%
\sqrt{\varepsilon ^{2}-m^{2}}\;\frac{1}{\exp \left[ \frac{1}{T}\left(
\varepsilon -\mu \right) \right] +1}\qquad ,\qquad \qquad \qquad \,(a)
\label{1}
\end{equation}
\begin{equation*}
\frac{E}{V}=\frac{3}{\pi ^{2}}\int_{m}^{\infty }d\varepsilon
\;\varepsilon ^{2}\;\sqrt{\varepsilon ^{2}-m^{2}}\;\frac{1}{\exp
\left[ \frac{1}{T}\left( \varepsilon -\mu \right) \right]
+1}\qquad ,\,\qquad \qquad \;(b)
\end{equation*}
\begin{equation*}
p=\frac{1}{\pi ^{2}}\int_{m}^{\infty }d\varepsilon \;\left(
\varepsilon ^{2}-m^{2}\right) ^{\frac{3}{2}}\;\frac{1}{\exp \left[
\frac{1}{T}\left( \varepsilon -\mu \right) \right] +1}\qquad
\,.\qquad \qquad \;\;\;\;(c)
\end{equation*}

Expanding in powers of $\frac{m^{2}}{\varepsilon ^{2}}$ and calling $a=\frac{%
m}{T}\,,\;x=\frac{\mu }{T}\,,$%
\begin{equation}
\frac{\pi ^{2}}{3T^{3}}\,\frac{N}{V}\simeq \exp x\int_{\exp a}^{\exp \Lambda
}\frac{d\eta }{\eta \left( \eta +\exp x\right) }\left( \ln ^{2}\eta -\frac{%
a^{2}}{2}-\frac{a^{4}}{8}\,\frac{1}{\ln ^{2}\eta }\right) \qquad
,\;\;(a) \label{2}
\end{equation}
\begin{equation*}
\frac{\pi ^{2}}{3T^{4}}\,\frac{E}{V}\simeq \exp x\int_{\exp a}^{\exp \Lambda
}\frac{d\eta }{\eta \left( \eta +\exp x\right) }\left( \ln ^{3}\eta -\frac{%
a^{2}}{2}\ln \eta -\frac{a^{4}}{8}\,\frac{1}{\ln \eta }\right)
,\;(b)
\end{equation*}
\begin{equation*}
\frac{\pi ^{2}}{T^{4}}\,p\simeq \exp x\int_{\exp a}^{\exp \Lambda }\frac{%
d\eta }{\eta \left( \eta +\exp x\right) }\left( \ln ^{3}\eta -\frac{3}{2}%
a^{2}\ln \eta +\frac{3}{8}\,\frac{a^{4}}{\ln \eta
}\right)\;,\;\;(c)
\end{equation*}
in terms of a cut-off $\Lambda $ which will not appear in the final results.

We will consider quarks $u$ and $d$ with $a\simeq 0$ and quark $s$ with $%
a\simeq 1$ at the confinement transition $T$.

If $x<a$ , which will happen for $s$ at the beginning of bubble contraction
and for all antiquarks because $\mu _{\overline{q}}<0\;,$ we may expand the
denominators in powers of $\frac{\exp x}{\eta }$ to obtain
\begin{eqnarray}
\frac{\pi ^{2}}{3T^{3}}\;\frac{N}{V}&\simeq&\exp (x-a)
\;\left(2+2a+\frac{a^{2}}{2}-\frac{a^{3}}{8}\right)-\qquad\qquad\qquad\qquad
\;\;\;\;(a)\label{3}\\
&&-\frac{\exp[2(x-a)]}{2}\left(\frac{1}{2}+a+\frac{a^{2}}{2}-\frac{a^{3}}{4}\right)+\nonumber\\
&&+\frac{\exp[3(x-a)]}{3}\left(\frac{2}{9}+\frac{2}{3}a+\frac{a^{2}}{2}-\frac{3}{8}a^{3}\right)-\nonumber\\
&&-\frac{a^{4}}{8}[\exp(x) Ei(-a)-2\exp(2x) Ei(-2a)+3\exp(3x)
Ei(-3a)]\;,\nonumber
\end{eqnarray}
\begin{eqnarray*}
\frac{\pi ^{2}}{3T^{4}}\;\frac{E}{V} &\simeq &\exp \left(
x-a\right) \;\left( 6+6a+\frac{5}{2}a^{2}+\frac{a^{3}}{2}\right)
-\qquad \qquad \qquad \;\;\;\;\;\;\;\;\left( b\right) \\
&&-\frac{\exp \left[ 2\left( x-a\right) \right] }{2}\left(
\frac{3}{4}+\frac{3}{2}a+\frac{5}{4}a^{2}+\frac{a^{3}}{2}\right) +
\\
&&+\frac{\exp \left[ 3\left( x-a\right) \right] }{3}\left(
\frac{2}{9}+\frac{2}{3}a+\frac{5}{6}a^{2}+\frac{a^{3}}{2}\right) +
\\ &&+\frac{a^{4}}{8}\left[ \exp(x)\;Ei\left( -a\right) -
\exp \left(2x\right) Ei(-2a)+\exp \left( 3x\right)Ei\left(
-3a\right)\right] \;,
\end{eqnarray*}
\begin{eqnarray*}
\frac{\pi ^{2}}{T^{4}}p &\simeq &\exp \left( x-a\right) \;\left(
6+6a+\frac{3}{2}a^{2}-\frac{a^{3}}{2}\right) -\qquad \qquad \qquad
\qquad \,\;\;\;\left( c\right) \\ &&-\frac{\exp \left[ 2\left(
x-a\right) \right] }{2}\left(
\frac{3}{4}+\frac{3}{2}a+\frac{3}{4}a^{2}-\frac{a^{3}}{2}\right) +
\\
&&+\frac{\exp \left[ 3\left( x-a\right) \right] }{3}\left(
\frac{2}{9}+\frac{2}{3}a+\frac{a^{2}}{2}-\frac{a^{3}}{2}\right) -
\\
&&-\frac{3}{8}a^{4}\left[ \exp(x)\;Ei\left( -a\right) -\exp
\left(2x\right) Ei(-2a)+\exp \left( 3x\right) Ei\left( -3a\right)
\right] \;.
\end{eqnarray*}

On the other hand for $x>a$ , which is needed for $u$ and $d$ and occurs for
$s$ at the end of the bubble contraction, the integrals of Eq.(2) will be
the sum of two parts one of the same type but with lower limit $\exp x$ to
allow the expansion in powers of $\frac{\exp x}{\eta }$ and another between $%
\exp a$ and $\exp x$ where the expansion of the denominator is in powers of $%
\frac{\eta }{\exp x}$ . The results are
\begin{eqnarray}
\frac{\pi ^{2}}{3T^{3}}\;\frac{N}{V} &\simeq &\frac{x^{3}}{3}+\frac{x^{2}}{3}%
+\frac{29}{9}x+\frac{2}{27}-\frac{1}{2}a^{2}x-\qquad \qquad \qquad
\qquad \qquad \;\left( a\right)  \label{4} \\
&&-\frac{1}{6}a^{2}+\frac{a^{3}}{24}+\exp \left[ -\left(
x-a\right) \right] \left(
2-2a+\frac{a^{2}}{2}+\frac{a^{3}}{8}\right) -  \notag \\
&&-\frac{\exp[ -2( x-a)}{2}\left( \frac{1}{2}-a+
\frac{a^{2}}{2}+\frac{a^{3}}{4}\right) +\frac{a^{4}}{8}[-\exp
(x)\,Ei\left( -x\right) +  \notag \\ &&+2\exp (2x)\,Ei(-2x)-3\exp
\left( 3x\right) \,Ei\left(-3x\right) +\exp \left( -x\right)
\,Ei\left( x\right) -  \notag \\ &&-2\exp \left( -2x\right)
\,Ei\left( 2x\right) -\exp \left( -x\right) \,Ei\left( a\right)
+2\exp \left( -2x\right) \,Ei\left( 2a\right) ]\;,  \notag
\end{eqnarray}
\begin{eqnarray*}
\frac{\pi ^{2}}{3T^{4}}\;\frac{E}{V} &\simeq &\frac{x^{4}}{4}+\frac{x^{3}}{3}%
+\frac{29}{6}x^{2}+\frac{2}{9}x+\frac{1223}{108}-\frac{1}{4}%
a^{2}x^{2}-\qquad \qquad \;\;\;\;\;\,\left( b\right) \\
&&-\frac{1}{6}a^{2}x-\frac{29}{36}a^{2}+\exp \left[ -\left( x-a\right) %
\right] \left( -6+6a-\frac{5}{2}a^{2}+\frac{a^{3}}{2}\right) + \\
&&+\frac{\exp \left[ -2\left( x-a\right) \right] }{2}\left( \frac{3}{4}-%
\frac{3}{2}a+\frac{5}{4}a^{2}-\frac{a^{3}}{2}\right)
+\frac{a^{4}}{8}[\exp(x)\,Ei\left( -x\right) - \\ &&-\exp
(2x)\,Ei(-2x)+\exp \left( 3x\right) \,Ei\left(-3x\right) +\exp
\left( -x\right) \,Ei\left( x\right) - \\ &&-\exp \left(
-2x\right) \,Ei\left( 2x\right) -\exp \left( -x\right)\,Ei\left(
a\right) + \\ &&+\exp \left( -2x\right) \,Ei\left( 2a\right) -\ln
x+\ln a]\;,
\end{eqnarray*}
\begin{eqnarray*}
\frac{\pi ^{2}}{T^{4}}p &\simeq
&\frac{x^{4}}{4}+\frac{x^{3}}{3}+\frac{29}{6}
x^{2}+\frac{2}{9}x+\frac{1223}{108}+\frac{a^{4}}{2}-\qquad \qquad
\qquad \qquad \;\;\left( c\right) \\ &&-\frac{3}{2}a^{2}\left(
\frac{x^{2}}{2}+\frac{x}{3}+\frac{29}{18}\right) +\exp \left[
-\left( x-a\right) \right] \left(
-6+6a-\frac{3}{2}a^{2}-\frac{a^{3}}{2}\right) + \\ &&+\frac{\exp
\left[ -2\left( x-a\right) \right] }{2}\left( \frac{3}{4}-
\frac{3}{2}a+\frac{3}{4}a^{2}+\frac{a^{3}}{2}\right)
+\frac{3}{8}a^{4}[-\exp(x)\,Ei\left( -x\right) + \\ &&+\exp
(2x)\,Ei(-2x)-\exp \left( 3x\right) \,Ei\left(-3x\right) -\exp
\left( -x\right) \,Ei\left( x\right) + \\ &&+\exp \left(
-2x\right) \,Ei\left( 2x\right) +\exp \left( -x\right) \,Ei\left(
a\right) - \\ &&-\exp \left( -2x\right) \,Ei\left( 2a\right) +\ln
x-\ln a]\;.
\end{eqnarray*}

To analyze the relative stability of states of systems with variable number
of particles we must compare the free energy
\begin{equation}
\frac{J}{V}=\frac{E}{V}-T\frac{S}{V}-\mu \frac{N}{V}=-p\qquad ,  \label{5}
\end{equation}
summed over quarks and antiquarks of all flavours.

We now turn to the possibility that quarks form Cooper pairs.

For the case where $\mu >>m_{s}$ and $T$ is not too high it seems clear that
di-quark condensates are formed due to their interaction which is attractive
in the colour antisymmetric multiplet $\overline{3}$, favoured in the
antisymmetric zero total spin and flavour antisymmetry to allow total
antisymmetry of fermions, \textit{i.e.}
\begin{equation}
\left\langle q_{iL}^{\alpha }\left( \mathbf{p}\right) q_{iL}^{\beta }\left( -%
\mathbf{p}\right) \right\rangle =\Delta \sum\limits_{c=1}^{3}\varepsilon
^{\alpha \beta c}\varepsilon _{ijc}\qquad ,  \label{6}
\end{equation}
where $\alpha $ and $i$ are respectively colour and flavour indices and the
pair has the same chirality because the momenta of the partners are equal
and opposite. Analogous expression holds for $R$ chirality.

In this $CFL$ phase both the gauge $SU(3)_{c}$ and the approximate global
chiral $SU(3)_{L,R}$ symmetry are broken giving respectively massive gluons
and mesons as Goldstone particles.

The gap $\Delta $ can be calculated in the Nambu-Jona Lasinio approximation
of $QCD$ where an effective 4-fermion interaction with coupling $G$ is
considered.

For $T=0$ the estimation\cite{8}
\begin{equation}
\Delta _{0}\simeq \mu \,\exp \left( \frac{-\pi ^{2}}{2G\mu ^{2}}\right)
\label{7}
\end{equation}
may be given with $G\simeq \frac{1}{GeV^{2}}\;.$

More precise evaluations\cite{2} are done with perturbative $QCD$ with
coupling $g$ giving
\begin{equation}
\Delta _{0}=b\,\,\frac{\mu }{g^{5}}\exp \left( -\frac{3\,\pi ^{2}}{\sqrt{2}%
\,g}\right) \qquad ,  \label{8}
\end{equation}
where the coefficient is primarily $b=512\,\pi ^{4}\left( \frac{2}{N_{f}}%
\right) ^{\frac{5}{2}}$, $N_{f}$ is the number of flavours that we
take $3$ and $\alpha _{s}=\frac{g^{2}}{4\pi }\simeq \frac{0.7}{\ln
\left( \frac{\mu }{\Lambda _{QCD}}\right) }$ with $\Lambda
_{QCD}=200\,MeV.$ But additional corrections to $b$ may decrease
the above expression by a factor of $5$ or increase it by $20$.

For $\mu \sim 1\;GeV$ Eq.(8) gives $\Delta _{0}\sim 50\;MeV$ in agreement
with the result with Eq.(7). Even though one cannot trust the extrapolation
of the perturbative calculation down to $\sim 1\,GeV$ , Eq.(8) shows clearly
that because of the asymptotic freedom limit $g\rightarrow 0$ for $\mu
\rightarrow \infty $, finally $\Delta _{0}$ will decrease.

For $T>0$ there will be a non-vanishing gap provided $T<T_{c}\simeq
0.57\Delta _{0}$. Therefore for very large $\mu $ colour superconductivity
might be achieved even at the confinement temperature $T\simeq 150\,MeV$.

To evaluate the free energy $J$ at $T=0$ one has to sum the expression for
the unpaired case but with the restriction of equal Fermi momentum for all
flavours, in order to have the same number of quarks, plus a contribution
coming from the gap\cite{9}, \textit{i.e.}
\begin{equation}
\left. \frac{J}{V}\right| _{paired}^{(T=0)}=\left. \frac{J}{V}\right|
_{unpair}^{(T=0)}\left( p_{F}^{u}=p_{F}^{d}=p_{F}^{s}\right) -\frac{3}{\pi
^{2}}\Delta _{0}^{2}\mu ^{2}\quad .  \label{9}
\end{equation}

The condition of equal Fermi momenta increases $J_{unpair}$ compared to the
same expression for free quarks but the difference is compensated by the
last term of Eq.(9) for large enough $\mu $.

For $T>0$ the corresponding formula will have $J_{unpair}$ with chemical
potentials such that the densities of quarks of different flavours are equal
\cite{10} and a contribution of gap which vanishes for a continuous
transition\cite{8} at $T=T_{c}$ , \textit{i.e.}
\begin{equation}
\left. \frac{J}{V}\right| _{paired}=\left. \frac{J}{V}\right|
_{unpair}\left( n_{u}=n_{d}=n_{s}\right) -\frac{18\mu ^{2}T_{c}^{2}}{7\zeta
\left( 3\right) }\left( 1-\frac{T}{T_{c}}\right) ^{2}\quad .  \label{10}
\end{equation}

The pressure of the paired quarks will be therefore larger than that of
non-interacting quarks for $T$ clearly below $T_{c}$.

\bigskip

\section{Calculation of bubble contraction}

We will consider the force applied to the bubble, accepting that the
confinement transition was first order, causing the increase of its
relativistic momentum in radial direction together with release of the
momentum of neutrinos emitted due to quark-antiquark annihilation
\begin{equation}
F=\frac{d}{dt}\left( \frac{M}{\sqrt{1-v^{2}}}\;v\right) +\frac{dP_{\nu }}{dt}%
\qquad ,  \label{11}
\end{equation}
where the rest mass $M$ comes from quark, antiquark and false vacuum energy
contributions. The force will result from the difference $\Delta p$ between
the pressure of false vacuum and that of quarks and antiquarks inside the
bubble applied to its surface.

Taking the bubble radius $r$ negative in order to have positive velocity $v$
during contraction, from Eq.(11) we have
\begin{equation}
\Delta p\,4\pi \,r^{2}\sqrt{1-y}=y\,\frac{dM}{dr}+\frac{M}{1-y}\,\frac{1}{2}%
\,\frac{dy}{dr}+\frac{dP_{\nu }}{dt}\sqrt{1-y}\qquad ,  \label{12}
\end{equation}
where $y=v^{2}$.

In the first stage of the bubble compression $\Delta p>0$ and the rest mass
decreases because quarks and antiquarks annihilate to mantain the density
determined by thermodynamic equilibrium. On its side the momentum of the
copiously produced neutrinos, as we will see, tends to compensate the
decrease of the first term of $RHS$ of Eq.(12). Therefore, the second term
shows an increase of the velocity even though smaller than without the $\nu $
term.

On the other hand, when the bubble radius becomes small, almost all the
antiquarks have been annihilated and the produced neutrinos momentum is
negligible whereas the quark chemical potential raises rapidly the pressure
so that $\Delta p<0$ and gives also an increase of $M$. As a consequence the
velocity decreases and finally the bubble compression stops.

In our calculation we consider the system with neutral electric charge and
make the approximations of neglecting the presence of electrons and of
surface effects.

The condition of zero charge density gives the baryonic density as
\begin{equation}
\frac{N_{B}}{V}=\frac{N_{u}-N_{\overline{u}}}{V}=\frac{1}{2}\left( \frac{%
N_{d}-N_{\overline{d}}}{V}+\frac{N_{s}-N_{\overline{s}}}{V}\right) \qquad .
\label{13}
\end{equation}

At the beginning $\mu <<T$ because with this approximation Eq.(1a) gives
\begin{equation}
\frac{N_{q}}{V}\simeq \frac{5.46}{\pi ^{2}}T^{3}+\frac{4.92}{\pi ^{2}}%
T^{2}\mu \qquad ,  \label{14}
\end{equation}
so that the contribution of quarks to baryonic density is
\begin{equation}
\frac{N_{u}-N_{\overline{u}}}{V}\simeq \frac{9.84}{\pi ^{2}}T^{2}\mu \qquad .
\label{15}
\end{equation}

Since the entropy density is $s\sim T^{3}$, the normalized matter-antimatter
asymmetry necessary for the primordial nucleosynthesis gives
\begin{equation}
\frac{n_{B}}{s}\sim 10^{-10}\sim \frac{\mu }{T}\qquad ,  \label{16}
\end{equation}
determining for the transition temperature $\sim 100\;MeV$ the consistently
small value of chemical potential $\mu \sim 10^{-2}eV.$

If the bubble suffers no evaporation it will conserve the baryonic number
during the contraction so that, being\cite{6} the initial radius $\sim 1\;cm$%
, according to Eq.(15) $\mu $ will increase reaching $T$ for a radius $\sim
10^{-3}\;cm.$

Initially the bubble will be compressed by an ''external pressure'' given by
the negative of the energy density of the false vacuum $\rho _{V}\simeq
\Lambda _{QCD}^{4}.$ But when $\mu $ surmounts $T$ and practically no more
antiquarks survive, according to Eq.(4c) the pressure of the 3 flavours may
be estimated by $\sim \frac{3}{4\pi ^{2}}\mu ^{4}$ which will overcome the
external one and $\Delta p$ of Eq.(12) will change its sign.

Similarly, at the beginning when $\mu $ is very small the rest mass of the
bubble is given by an almost constant energy density according $to$
Eqs.(3b,4b) plus the constant false vacuum energy density $\Lambda
_{QCD}^{4} $, so that $M$ will decrease with $V$. But when $\mu >T$ the term
$\sim \mu ^{4}$ in the energy density will dominate making $M$ increase with
the bubble contraction.

Finally, taking the effective theory as indicative for the values of
chemical potential achieved in the process, according to Eq.(7) when $\mu
\stackrel{>}{\sim} 10\,T$ the gap $\Delta _{0}$ may be sufficiently large to allow $%
T<T_{c}$ and make the bubble colour superconducting before it stops.

The numerical procedure for the evaluation of Eq.(12) has been as follows.
We started with a radius variable $r_{0}=-1\;cm$ and zero velocity $v_{0}=0.$
For the different steps the conservation of baryonic number $N_{B}$ gave
through Eq.(13) the chemical potential $\mu _{u}$ and a relation between $%
\mu _{d}$ and $\mu _{s}$. These two latter chemical potentials
were determined by minimizing the energy at each radius. Then the
pressure given by quarks and antiquarks could be calculated. The
precise value of the negative pressure of false vacuum was
adjusted to be slightly larger than that of quarks and antiquarks
in the initial stage in order to have the minimal compression
condition.

\begin{figure}[t]
\setlength{\unitlength}{1cm} \vspace{7cm}
\includegraphics{C:/users/milva/a1.eps}
\caption{\small Rest mass $M$ of bubble including quark-antiquark
thermodynamic \hspace{1cm} contributions at $T=150\, MeV$ and
false vacuum energy.}
\end{figure}

\begin{figure}[h]
\setlength{\unitlength}{1cm} \vspace{7cm}
\includegraphics{C:/users/milva/a2.eps}
\caption{\small Bubble chemical potential and internal pressure.
Continuous line $\log \mu _{u}[ MeV]$, broken line $\log(
\frac{\pi ^{2}p}{T^{4}}).$}
\end{figure}

\begin{figure}[t]
\setlength{\unitlength}{1cm} \vspace{7cm}
\includegraphics{C:/users/milva/a3.eps}
\caption{\small Velocity of contraction of surviving quark matter
bubble at $T=150\,MeV.$}
\end{figure}

Regarding the inclusion of the momentum of neutrinos in Eq.(11) it
is convenient to consider the two stages of $\mu <<T$ and
$\mu\stackrel{>}{\sim}T$.

In the former we may approximate the energy density as a constant $\rho
=\rho _{V}+\rho _{q,\overline{q}}$ where $\rho _{q,\overline{q}}$ is due to
all quarks and antiquarks, so that $\frac{dM}{dr}=-\rho \;4\pi \;r^{2}$. In
the rest frame $\rho _{q,\overline{q}}$ is transformed into neutrinos which
will be isotropically distributed with an average momentum $k$ such that
\begin{equation}
dE_{\nu }=\rho _{q,\overline{q}}\,4\,\pi \,\,r^{2}\,v\,dt=4\,\pi
\,k^{2}\sigma \qquad ,  \label{17}
\end{equation}
where $\sigma $ is surface energy density in space $\mathbf{k}.$

Taking the motion in the direction 3, the momentum seen from lab frame will
be $k_{3}^{\prime }=\frac{k\cos \theta +vk}{\sqrt{1-v^{2}}}$ . Therefore
there will be a total momentum in lab system
\begin{equation}
dP_{3}=k^{2}\int \left( \frac{\sigma }{k}\right) \,k_{3}^{\prime }\,\sin
\theta \,d\theta \,d\varphi =\frac{v^{2}}{\sqrt{1-v^{2}}}\,\rho _{q,%
\overline{q}}\,4\,\pi \,r^{2}\,dt\qquad ,  \label{18}
\end{equation}
and the neutrino momentum will cancel the contribution of $\rho _{q,%
\overline{q}}$ in $\frac{dM}{dt}$ of Eq.(12), leaving
\begin{equation}
\Delta p\sqrt{1-y}=-\rho _{V}\,y+\frac{1}{6}\,\left| r\right| \,\rho \,\frac{%
1}{1-y}\,\frac{dy}{dr}\qquad .  \label{19}
\end{equation}

In the small radius stage when $x\stackrel{>}{\sim}1$, the
antiquarks have been almost
all annihilated so that $\rho _{\overline{q}}<<\rho _{q}$ and only $2\rho _{%
\overline{q}}$ will transform into neutrinos. Therefore Eq.(12) must be used
with
\begin{equation}
\sqrt{1-y}\,\frac{dP_{\nu }}{dt}=4\,\pi \,r^{2}\,y\,2\rho _{\overline{q}%
}\qquad .  \label{20}
\end{equation}

In this last part of bubble compression, when $T<T_{c}$ the chemical
potentials $\mu _{d}$ and $\mu _{s}$ will be determined by the condition of
having equal density of $u$, $d$ and $s$. The pressure of quark matter will
be increased by the gap term of Eq.(10).

In this way the properties of the bubble were calculated,
\textit{i.e.} the rest mass $M$ (Fig.1) the chemical potential and
the internal pressure $p$ (Fig.2) and the velocity of contraction
$v$ (Fig.3).

From them it is seen that the bubble stops its compression for a
final radius $R_{f}$ slightly below $10^{-4}cm$, its rest mass
starts to increase for radius slightly below $10^{-3}cm$ when also
the increase of the internal pressure overcomes the external one,
and in between these sizes $\mu $ reaches a value $\sim
1.5\,GeV\;$so that $T<T_{c}$ allowing the possibility that the
bubble quark matter becomes colour superconducting.

Finally, the interval of time for the bubble contraction
\begin{equation}
\Delta t=\int_{r_{0}}^{r_{f}}\frac{dr}{v}\qquad ,  \label{21}
\end{equation}
can be calculated numerically giving $\Delta t\sim 10^{-10}\sec $ much
smaller than $10^{-5}\sec $ so that the approximation of $T\sim $ constant
is consistent.

\section{Conclusions}

According to our model, the bubbles of quark matter surviving the
confinement transition may reach such high values of chemical potential,
larger than those in neutron stars, that could enter the colour
superconducting $CFL$ phase before their contraction stops becoming
therefore stable. A more accurate analysis of the order of the transition
between $QGP$ and $CFL$ phases\cite{11} should be done to establish with
certainty the dependence of $T_{c}$ on $\mu .$

There are several physical considerations that may affect the above result.
One is the emission of neutrons during the bubble compression which would
make the increase of chemical potential softer. Another thing is the
possible inclusion of a density of electrons which would alter the quark
chemical potentials to satisfy charge neutrality. Finally, taking into
account surface effects the condition of charge neutrality of the bubble may
not be realistic.

In case that the bubbles, perhaps smaller, can nevertheless become colour
superconducting at the relatively large temperature $\sim 150\,MeV$, one has
to take into account that if their energy per baryon is larger than $%
940\,MeV $ they will emit neutrons during the cooling till the
present time. If this process leaves bubbles of atomic number
$A>10^{3}$ at $T\sim 0$ they would be absolutely stable\cite{12}
and might be detected as strangelets by high-altitude
observatories since the collisions in the upper part of the
atmosphere would not reduce substantially their mass maintaining
therefore their stability.

\vspace{1cm}

\textbf{Acknowledgements.}We have been stimulated by interesting
talks with Adriano Di Giacomo at the Blokhintsev Conference of
Dubna and with Vladimir Nikolaev at the Nuclear Theory Workshop of
Rila.

\end{document}